\documentclass[twocolumn]{aastex62}
\usepackage{graphicx}
\usepackage[space]{grffile}
\usepackage{latexsym}
\usepackage{amsfonts,amsmath,amssymb}
\usepackage{url}
\usepackage[utf8]{inputenc}
\usepackage{fancyref}
\usepackage{lipsum}
\usepackage{ulem} 
\usepackage{color} 
\usepackage{comment}
\usepackage{dcolumn}
\usepackage{multirow}
\usepackage{bm}

\revised{\today}

\defcitealias{Gould2013}{G13}
\defcitealias{Bachelet+2018}{B18}
\defcitealias{Mogavero+2016}{M16}
\defcitealias{Penny+2019}{P19}
\defcitealias{Cumming+2008}{C08}

\shorttitle{{\it Roman} Xallarap Planets}
\shortauthors{Miyazaki et al.}
\begin{document}

\title{Revealing Short-period Exoplanets and Brown Dwarfs in the Galactic Bulge using the Microlensing Xallarap Effect with the {\it Nancy Grace Roman Space Telescope}}

\correspondingauthor{Shota Miyazaki}
\author[0000-0001-9818-1513]{Shota Miyazaki$^{\dag}$}
\affil{Department of Earth and Space Science, Graduate School of Science, Osaka University, 1-1 Machikaneyama, Toyonaka, Osaka 560-0043, Japan}
\author[0000-0001-9397-4768]{Samson A. Johnson}
\affil{Department of Astronomy, The Ohio State University, 140 West 18th Avenue, Columbus OH 43210, USA}
\author{Takahiro Sumi}
\affil{Department of Earth and Space Science, Graduate School of Science, Osaka University, 1-1 Machikaneyama, Toyonaka, Osaka 560-0043, Japan}
\author[0000-0001-7506-5640]{Matthew T. Penny}
\affil{Department of Physics and Astronomy, Louisiana State University, Baton Rouge, LA 70803, USA}
\author[0000-0003-2302-9562]{Naoki Koshimoto}
\affil{Department of Astronomy, Graduate School of Science, The University of Tokyo, 7-3-1 Hongo, Bunkyo-ku, Tokyo 113-0033, Japan}
\author{Tsubasa Yamawaki}
\affil{Department of Earth and Space Science, Graduate School of Science, Osaka University, 1-1 Machikaneyama, Toyonaka, Osaka 560-0043, Japan}

\begin{abstract}
The {\it Nancy Grace Roman Space Telescope} ({\it Roman}) will provide an enormous number of microlensing light curves with much better photometric precisions than ongoing ground-based observations. 
Such light curves will enable us to observe high-order microlensing effects which have been previously difficult to detect.
In this paper, we investigate {\it Roman}'s potential to detect and characterize short-period planets and brown dwarfs (BDs) in source systems using the orbital motion of source stars, the so-called xallarap effect.
We analytically estimate the measurement uncertainties of xallarap parameters using Fisher matrix analysis.
We show that the \textit{Roman} Galactic Exoplanet Survey (RGES) can detect warm Jupiters with masses down to 0.5 $M_{\rm Jup}$ and orbital periods of 30 days via the xallarap effect.
Assuming a planetary frequency function from \citet{Cumming+2008}, we find {\it Roman} will detect $\sim10$ hot and warm Jupiters and $\sim30$ close-in BDs around microlensed source stars during the microlensing survey.
These detections are likely to be accompanied by the measurements of the companion's masses and orbital elements, which will aid in the study of the physical properties for close-in planet and BD populations in the Galactic bulge. 
\end{abstract}

\section{Introduction} \label{sec:intro}
Gravitational microlensing \citep{Mao+1991,Bennett+1996} has a unique sensitivity to low-mass exoplanets beyond the snow line \citep{Hayashi+1985} where planet formation is considered active by the enhanced surface density of solid materials.
It has maximum sensitivity to planets (around the lens objects) with projected semi-major axes roughly equal to the projected Einstein ring radius $R_{\rm E}$, where
\begin{eqnarray}
R_{\rm E} = \left(\frac{4GM_L}{c^2}\frac{D_LD_{LS}}{D_S}\right)^{1/2}.
\end{eqnarray}
Here $D_S$ and $D_L$ are the distances of the source and lens from the Earth, $M_L$ is the mass of the lens, and $D_{LS}=D_S-D_L$.
For typical microlensing events toward the Galactic bulge ($D_S=8\;{\rm kpc},\;D_L=4\;{\rm kpc},\;M_L=0.3\;M_{\odot}$), $R_{\rm E}$ is $\sim2.3$ au.
Using this ``binary-lens'' channel of microlensing, the {\it Nancy Grace Roman Space Telescope} \citep[][previously named {\it WFIRST}, hereafter {\it Roman}]{Spergel+2015} will conduct the \textit{Roman} Galactic Exoplanet Survey and discover $\sim1400$ cold wide-orbit exoplanets \citep[][hereafter \citetalias{Penny+2019}]{Penny+2019} and provide an otherwise-inaccesible statistical sample of exoplanets in previously un-probed regions of exoplanet parameter space (see Figure 9 of \citetalias{Penny+2019}).

{\it Roman} will detect many thousands of microlensing light curves which will generally have better photometric precision than many ground-based microlensing surveys.
This will enable the measurement of high-order microlensing effects which have been previously difficult to detect.
One of the high-order effects that can be measurable in the {\it Roman} light curves is xallarap \citep{Griest+1992, Han+1997, Poindexter+2005}.
Xallarap is a microlensing effect where the reflex motion of a source star in a binary system modulates the magnification of the source star.
A more commonly known microlensing effect, orbital microlens parallax \citep{Gould2004}, also causes the variations with the same mechanism by the orbital motion of an observer.\footnote{Xallarap can be considered as the inverse of parallax and is a semordnilap.}
The xallarap amplitude $\xi_{\rm E}$ corresponds to the semi-major axis of the source star $a_S$ normalized by the angular Einstein radius $\theta_{\rm E}$ projected to the source plane, i.e., 
\begin{eqnarray} \label{eq:xa1}
\xi_{\rm E} = \frac{a_S}{D_S\theta_{\rm E}} = \frac{a_S}{\hat{r}_{\rm E}},
\end{eqnarray}
where $\hat{r}_{\rm E}$ is the projected Einstein radii.
We note that $a_S$ is the distance between the source and the center of masses of the source system.
Using Newton's version of Kepler's third law, we can derive following equations from the Equation (\ref{eq:xa1}),
\begin{eqnarray} \label{eq:xa2}
\xi_{\rm E} & = &  \frac{1\;{\rm au}}{\hat{r}_{\rm E}}\left(\frac{M_P}{M_\odot}\right)\left[\frac{M_\odot}{M_S+M_P}\frac{P_\xi}{1\;{\rm yr}}\right]^{2/3} \nonumber \\ 
					& \simeq & 2\times 10^{-5}\left(\frac{1\;{\rm au}}{D_S\theta_{\rm E}}\right)\left(\frac{M_P}{M_{\rm Jup}}\right) \left[\frac{M_\odot}{M_S+M_P}\frac{P_\xi}{1\;{\rm day}}\right]^{2/3} ,\\
M_S a_S  &=& M_P a_P \nonumber\\ 
\Rightarrow a &\equiv& a_S + a_P = \left(1+\frac{M_S}{M_P}\right)a_S ,
\end{eqnarray}
where $M_S$ and $M_P$ are masses of the source (host) and source companion, $P_{\xi}$ is the orbital period, $a_P$ is the semi-major axis of the source companion, and $a$ is the distance between the host and companion in the source system.

Equation (\ref{eq:xa2}) means that when a solar-type source star in the Galactic bulge $(D_S=8\;{\rm kpc})$ is accompanied by a planet with $M_P=10\;M_{\rm Jup}$ and $\;P_\xi=10\;{\rm days}$, the angular size of the semi-major axis of the source star orbit around the barycenter is a factor $10^{-4}$ smaller than $\theta_{\rm E}$. 
Present ground-based microlensing survey observations do not have typical sensitivities to detect such small fluctuations induced by planetary-mass source companions.

In several microlensing analyses, xallarap has been investigated to explain light curve deviations from a standard model \citep{Paczynski1986} which assumes uniform linear motions between the source, lens, and observers \citep[e.g.][]{Bennett+2008, Sumi+2016}.
However, identifying the xallarap signals clearly is rarely successful.
For example, \citet{Sumi+2010} analyzed a planetary microlensing event OGLE-2007-BLG-368 and found clear asymmetric features that can be interpreted as xallarap signals.
However, they could not conclude it because possible unknown systematics in the light curve could not be ruled out.
Recently, \citet{Miyazaki+2020} identified a significant xallarap signal in a planetary microlensing event OGLE-2013-BLG-0911.
Using the observed xallarap parameters, they concluded that there is a late M-dwarf orbiting the source star with a mass of $0.14^{+0.02}_{-0.02}\;M_{\odot}$ and an orbital period of $36.7^{+0.8}_{-0.7}\;{\rm days}$.
This is the first demonstration that dark, low-mass objects in the Galactic bulge can be detected and characterized via xallarap even with ground-based photometry. 
\citet{Rahvar+2009} suggested a possibility that planets orbiting sources in the Galactic bulge are detectable via xallarap with sufficiently good photometry.
With space-based photometry like {\it Roman}, planetary-mass objects might be detectable and characterizable via xallarap.

In this paper, we investigate the possibility of detecting planetary xallarap signals in the \textit{Roman} microlensing events. 
In Section \ref{sec:FMA}, we describe our Fisher matrix analysis and analytical quantification of  the ability of the {\it Roman} light curves to detect xallarap signals and characterize the physical properties of the source systems. 
To predict how many planets are detectable in the {\it Roman} mission via the xallarap effect, we apply our analysis to simulations of the \textit{Roman} survey in Section \ref{sec:LCS}.
Finally, we give our conclusion and discussion in Section \ref{sec:DIS}. 

\section{Fisher Matrix Analysis \label{sec:FMA}}
In this section, we conduct the Fisher matrix analysis based on the expected {\it Roman} observations and evaluate its sensitivity for xallarap.
\citet{Rahvar+2009} adopted the value of $\Delta\chi^2$ between the xallarap and non-xallarap (standard) models as the detection threshold of the source companion.
However, it could be insufficient for evaluating the ability to characterize the physical parameters of planets.
Further discussion on this is presented in Appendix \ref{sec:deltachi2}. 
The mechanisms of how xallarap affects light curves are essentially identical to the microlens parallax.
Therefore, we conduct the Fisher matrix analysis by modifying the formulas of parallax that are conducted by \citet{Gould2013}, \citet[][hereafter \citetalias{Mogavero+2016}]{Mogavero+2016}, and \citet[][hereafter \citetalias{Bachelet+2018}]{Bachelet+2018}.  

\subsection{Parameterization of the Xallarap effect}
\begin{figure}[ht]
\centering
\includegraphics[scale=0.19,bb= 6 6 1277 1076]{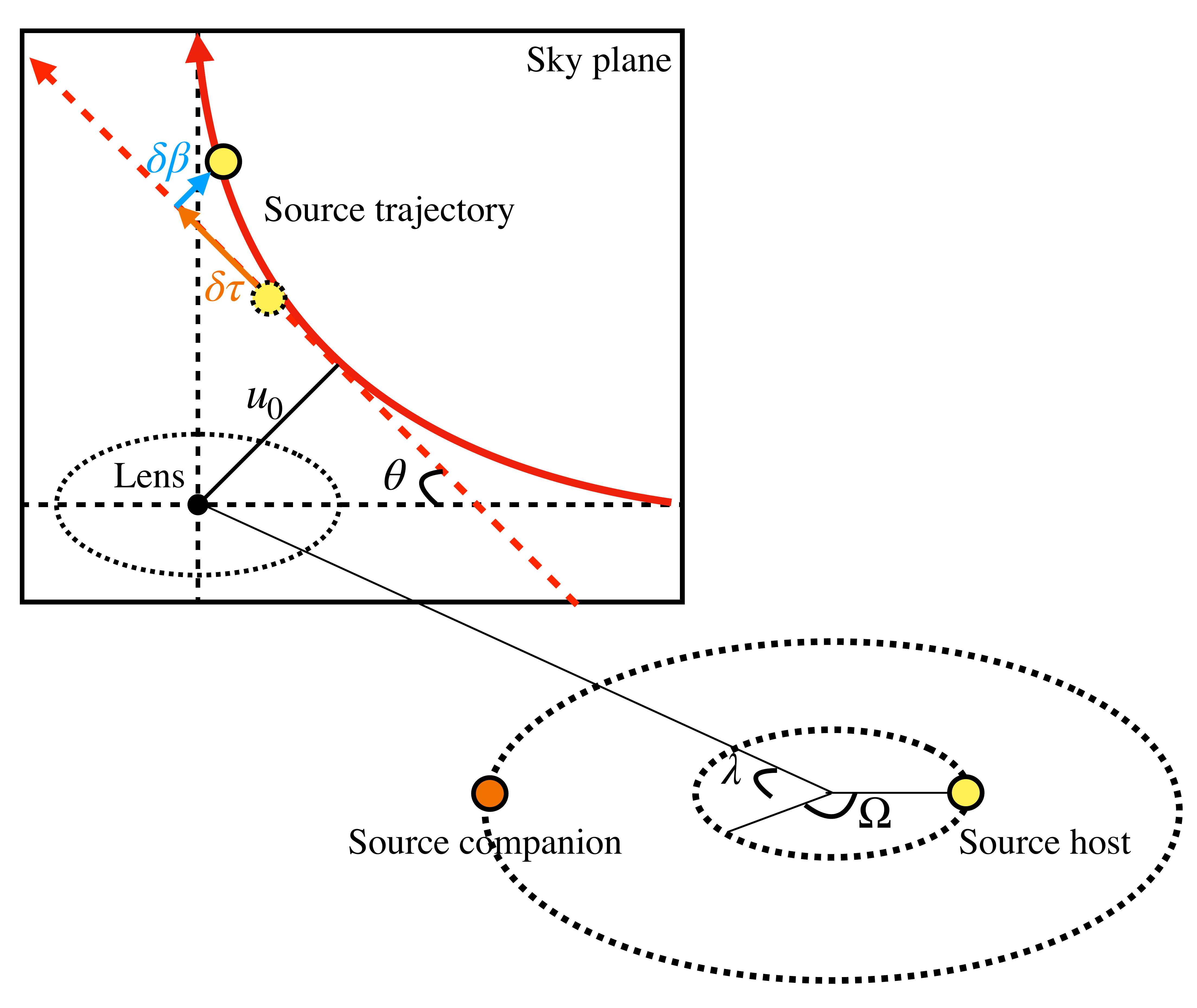}
\caption{
Schematic view of the xallarap problem.
Due to the source orbital motion, the source trajectory (solid red curve) deviates from the inertial trajectory (dashed red line). 
}
\label{fig:ponchie}
\end{figure}
\begin{figure*}[ht!]
\centering
\includegraphics[scale=0.5,bb= 0.000000 0.000000 944.318478 456.545000]{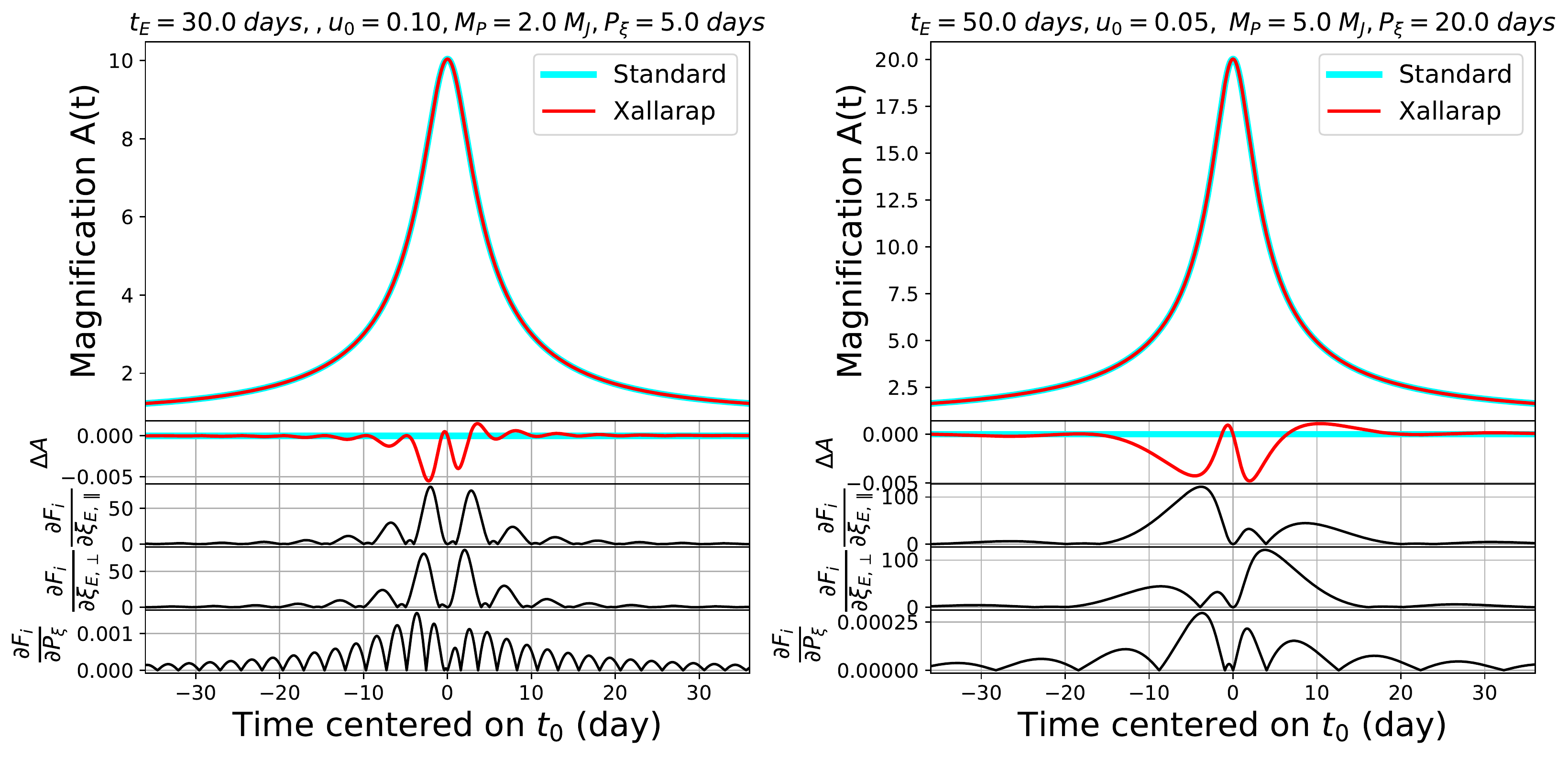}
\caption{
Two examples of simulated {\it Roman} light curves with xallarap.  
Top: The standard (cyan solid line) and xallarap (solid red line) light curves. 
Second to the Top: The residuals of the xallarap light curves relative to the standard ones.
The three bottom panels: The absolute values of components of integrands of the Fisher matrix $\partial F(t_k)/\partial \zeta_i$ at a given time $t_k$.
\label{fig:sample_lc}
}
\end{figure*}

Here we describe the xallarap effect observed by a single observatory. 
We follow \citetalias{Bachelet+2018}'s descriptions for the parallax effect observed by space-based observatories and then modify it for the case of xallarap effect.

In general, the observed flux of microlensing event is 
\begin{equation}
F=F_SA + F_B = \overline{F}\left[(1-\nu)A+\nu\right]
\end{equation}
where $F_S,\;F_B,\;\overline{F}(\equiv F_S+F_B)$ are the fluxes of the source, blend and baseline, respectively.
$\nu\;\equiv F_B/(F_S+F_B)$ denotes the blend flux ratio\footnote{$\overline{F}$ and $\nu$ are non-standard variables.}.
For a single-lens single-source (1L1S) model, the source flux magnification $A$ is described by
\begin{equation}
A(t)=\frac{u^2(t)+2}{u(t)\sqrt{u^2(t)+4}}, \label{eq:amp}
\end{equation}
where $u$ is the magnitude of the lens-source separation vector normalized by the angular Einstein radius $\theta_{\rm E}$, ${\bm u}$.
For uniform linear motions between the source, lens, and observers, $u(t)=\sqrt{\tau^2+u_0^2}$, where $\tau\equiv(t-t_0)/t_{\rm E}$, $t_0$ is the time of the magnification peak, and $u_0$ is the lens-source impact parameter normalized to $\theta_{\rm E}$.

Figure \ref{fig:ponchie} gives a schematic view of the xallarap problem.
Here we consider a planet in a circular orbit ($e=0$) around a source star with orbital period $P_{\xi}$ and mass $M_P$.
Then the source also orbits around the barycenter of the source system.
In this paper, we assume that the source companion contributes no flux to the event, i.e. it acts as a 1L1S event, not a binary source event \citep{Han+1998}.
The displacement of the source position due to the orbital motion can be described by
\begin{eqnarray}
{\bm S}(t) = \left(
\begin{array}{c}
s_1  \\
s_2 
\end{array}
\right)= \left(
\begin{array}{c}
\cos\Omega - \cos\phi_\xi \\
\sin\lambda_\xi(\sin\Omega - \sin\phi_\xi)
\end{array}
\right),
\end{eqnarray}
where $\Omega=\omega(t-t_0)+\phi_{\xi}$ and $\omega=2\pi/P_{\xi}$.
Here, $\lambda_{\xi}$ denotes the inclination of the source orbital plane with respect to the observer and $\phi_{\xi}$ denotes the orbital phase at $t_0$.
We define $\theta$ as the angle between the direction of the lens-source relative motion and the major axis of the source orbit projected on the sky. 
When we define the xallarap vector $\bm{\xi}_{\rm E}=(\xi_{{\rm E},\parallel}, \xi_{{\rm E},\perp})=\xi_{\rm E}(\cos\theta, \sin\theta)$, the displacement of the source position due to the orbit relative to the inertial source position is
\begin{eqnarray}
\delta\tau   & = & \bm{\xi}_{\rm E} \cdot {\bm S} \\
\delta\beta & = & \bm{\xi}_{\rm E}\times {\bm S} ,
\end{eqnarray}
where $|\bm{\xi}_E|$ = $a_S/(D_S\theta_{\rm E})$.
The lens-source separation vector ${\bm u}(t)$ can be described by 
\begin{eqnarray}
{\bm u}(t) = \left(
\begin{array}{c}
\tau^{\prime}\cos\theta -u^{\prime}\sin\theta\\
\tau^{\prime}\sin\theta +u^{\prime}\cos\theta
\end{array}
\right),
\end{eqnarray}
where $\tau^{\prime}=\tau+\delta\tau$ and $u^\prime=u_0+\delta\beta$.
The xallarap model can be described by ten parameters:
\begin{eqnarray}
{\bm \zeta}=(\overline{F}, \nu, t_0, t_{\rm E},u_0, \xi_{\rm E, \parallel}, \xi_{\rm E, \perp}, \phi_{\xi}, \lambda_{\xi}, P_{\xi}).
\end{eqnarray}

\subsection{Fisher Matrix Analysis} 
To estimate the expected uncertainty of each parameter $(\zeta_i)$ by the {\it Roman} microlensing survey, we calculate the Fisher matrix of the light curve model $F(t_k, {\bm \zeta})$ with given parameter set ${\bm \zeta}$.
Under the assumption of independent errors, the Fisher matrix elements $b_{i,j}$ can be written as 
 \begin{eqnarray}
b_{i,j} = \sum^N_{k=1} \frac{1}{\sigma_k^2} \frac{\partial F(t_k)}{\partial \zeta_{i}} \frac{\partial F(t_k)}{\partial \zeta_{j}} \label{eq:FM}
\end{eqnarray}
where $N$ is the total number of the data points, and $\sigma_k$ is the photometric error on data point at $t_k$.
Once the Fisher matrix is calculated, the covariance matrix for parameters ${\bm C}$ is given by its inverse matrix, i.e.,
\begin{eqnarray}
\bm{C} = \bm{b}^{-1}.
\end{eqnarray}
We follow the logic of \citetalias{Mogavero+2016} and discard (negligible) contribution of $\bar{F}$ from our Fisher matrix  analysis\footnote{An arbitrary uncertainty on $\overline{F}$ is achievable with a sufficient number of photometric observations while the event is at baseline. \textit{Roman} will collect $\sim40,000$ measurements in its \textit{W146} bandpass per light curve during the survey, which spans $6\times72$-d seasons spread out over 4.5 years.}.
In principle, the uncertainties for the xallarap amplitude depend on the event geometry, i.e., $\sigma^2_{\xi_{\rm E}}(\theta, \phi_{\xi})$.
To produce the results which are independent from the geometric conditions, \citetalias{Mogavero+2016} analytically found the minimum uncertainty on parallax measurement, $\sigma^2_{\pi_{\rm E},{\rm min}}$, which is independent of $\theta$.
We modify it for xallarap measurements $\sigma^2_{\xi_{\rm E},{\rm min}}(\phi_{\xi})$ as 
\begin{eqnarray}
\sigma^2_{\xi_{\rm E}, \pm}(\phi_\xi) &=& \frac{\sigma^2_{\xi_{\rm E,\parallel}} +\sigma^2_{\xi_{\rm E,\perp}}}{2} \nonumber \\
         &\pm& \frac{ \sqrt{(\sigma^2_{\xi_{\rm E,\parallel}} - \sigma^2_{\xi_{\rm E,\perp}})^2 + 4 {\rm cov}(\xi_{\rm E,\parallel}, \xi_{\rm E,\perp})^2}}{ 2 }, \nonumber\\
\sigma^2_{\xi_{\rm E},{\rm min}} &\equiv&\min_{\phi_\xi\in[0,2\pi]}\sigma^2_{\xi_{\rm E}, -}(\phi_\xi).
\end{eqnarray}
As \citetalias{Mogavero+2016} noted, for $P_{\xi}\ll u_0t_{\rm E}$, the covariance between the xallarap vector components $\xi_{\rm E, \parallel}$ and $\xi_{\rm E, \perp}$ disappear so that $\sigma_{\xi_{\rm E}}$ becomes independent of $\phi_{\xi}$.

\begin{figure*}[ht]
\centering
\includegraphics[scale=0.5,bb= 0.000000 0.000000 1011.034918 404.075625]{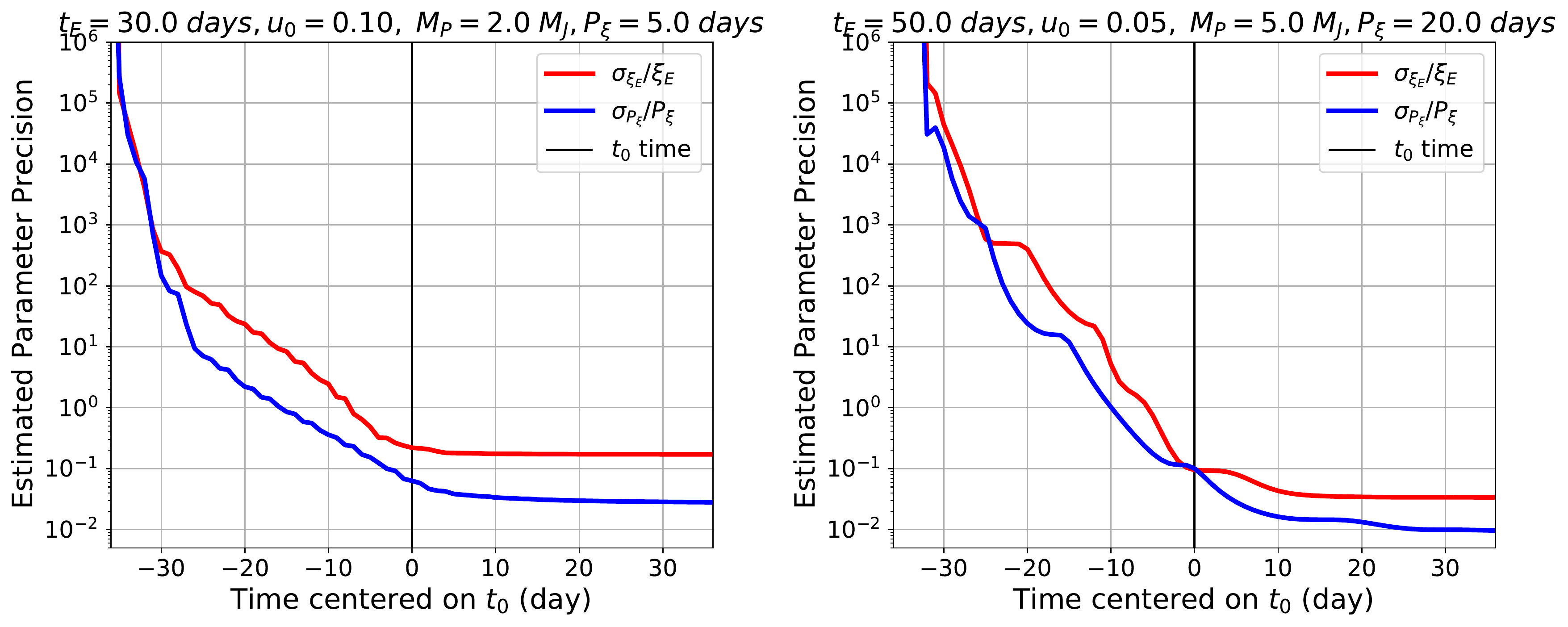}
\caption{
Parameter uncertainties for the xallarap amplitude $\xi_{\rm E}$ (red lines) and orbital period $P_{\xi}$ (blue lines) estimated at each time when a {\it Roman} observation is conducted.
To obtain these lines, we calculate the covariance matrices from the Fisher matrix at each observation.
The parameter conditions for the two panels are identical to that of Figure \ref{fig:sample_lc}.
\label{fig:cumu_sens}
}
\end{figure*}
\begin{figure*}[ht]
\centering
\includegraphics[scale=0.5,bb= 0.000000 0.000000 1011.034918 404.075625]{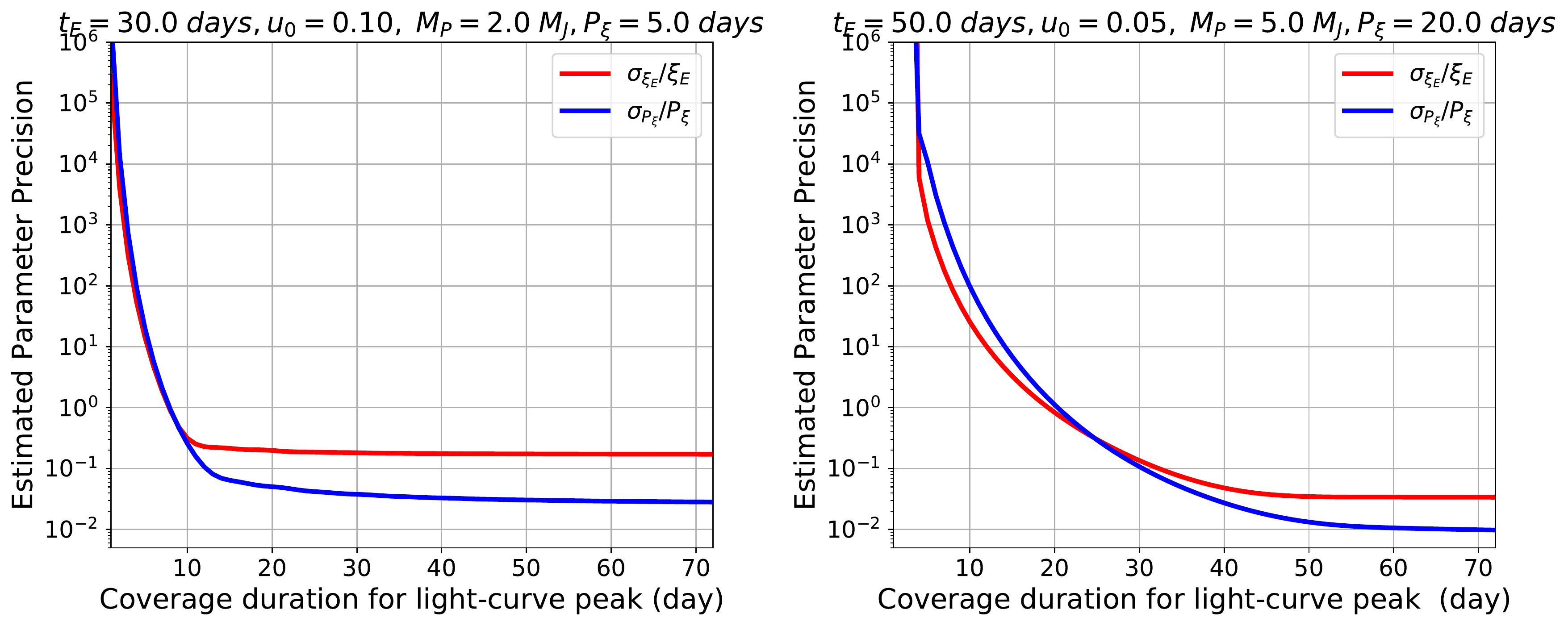}
\caption{
Parameter uncertainties for the xallarap amplitude $\xi_{\rm E}$ (red lines) and orbital period $P_{\xi}$ (blue lines) as a function of the coverage duration for the light-curve peak ($t_0$).
For plotting these, data points of \textit{Roman} light curve are uniformly distributed within the duration centered on $t_0$ and then we conduct the Fisher matrix analysis using the data points.
The parameter conditions for the two panels are identical to that of Figure \ref{fig:sample_lc}.
\label{fig:cumu_sens2}
}
\end{figure*}
\begin{figure*}
\centering
\includegraphics[scale=0.5,bb= 0.000000 0.000000 913.294375 739.545938]{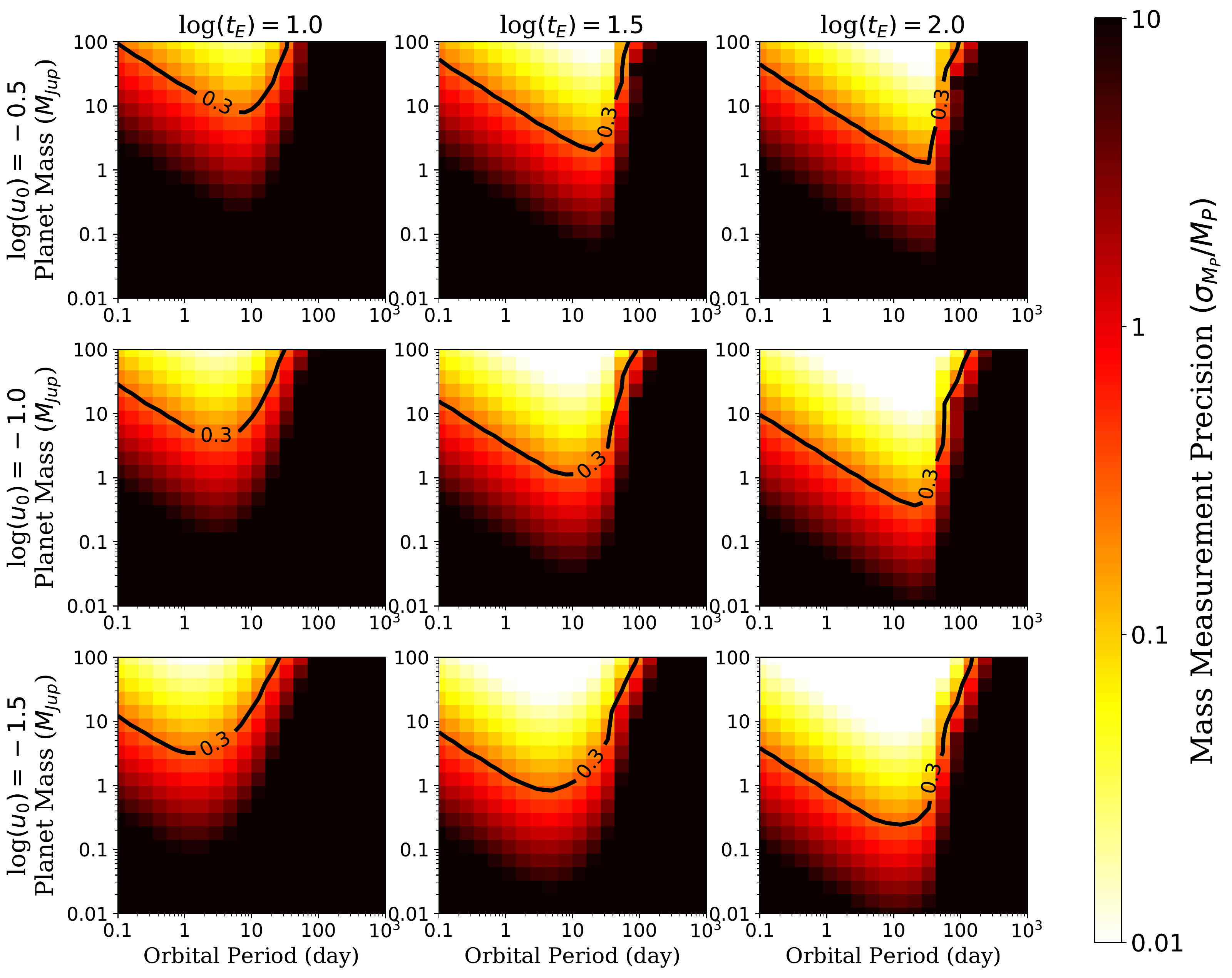}
\caption{
Color map distributions of the planetary mass error $\left(\frac{\sigma_{M_P}}{M_P}\right)$ for an event with the source magnitude of \textit{W146}$_S=18$ mag.
\label{fig:sens18}
}
\end{figure*} 
\begin{figure*}[ht!]
\centering
\includegraphics[scale=0.75,bb=0.000000 0.000000 639.547600 493.795125]{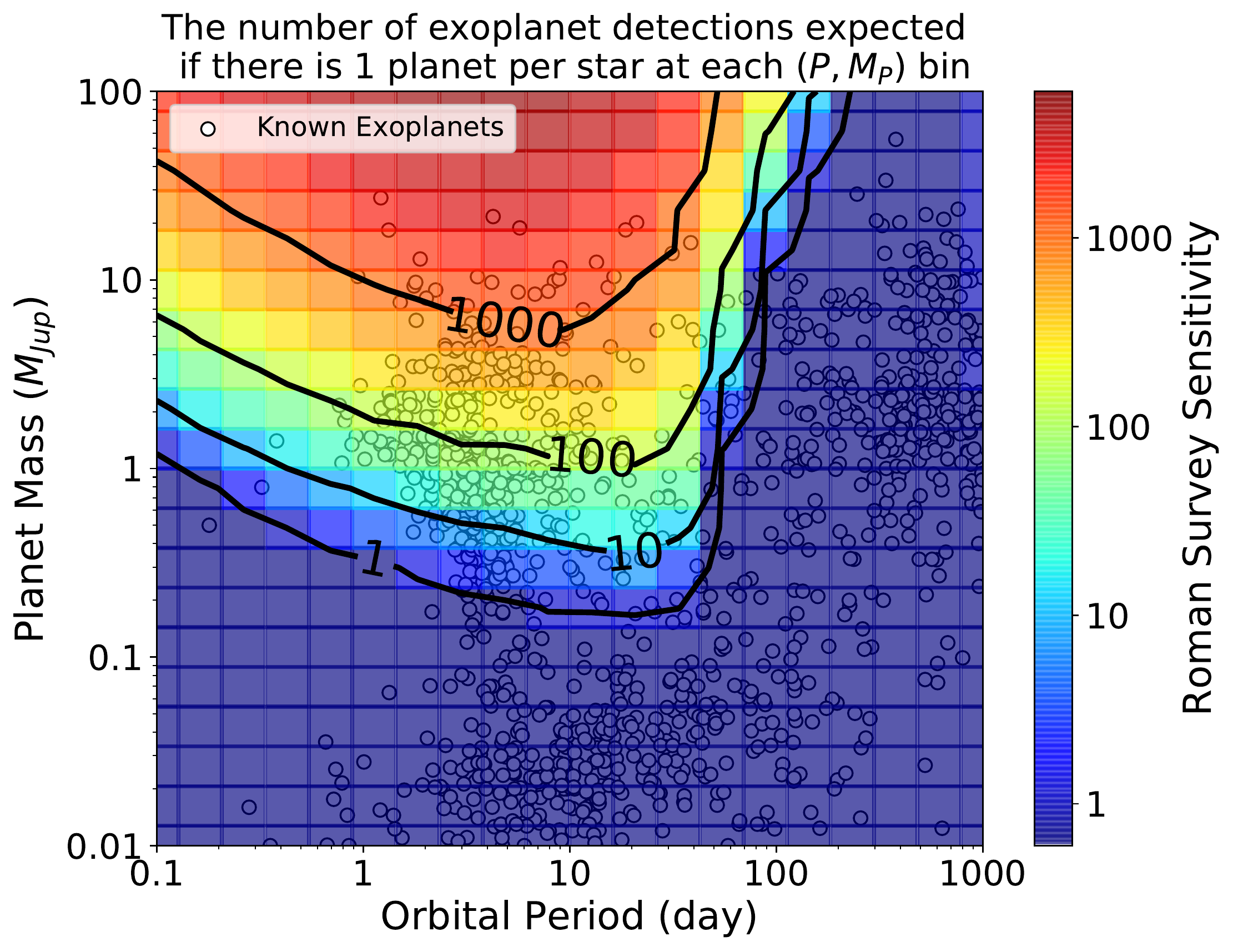}
\caption{
{\it Roman} xallarap sensitivity in the mass-orbital period plane.
The color map shows the number of detections for xallarap planets or brown dwarfs during the {\it Roman} mission at a given mass and orbital period grid if there is one planet per star at a given grid.
The open circles represent confirmed exoplanets referenced from \dataset[NASA Exoplanet Archive]{https://exoplanetarchive.ipac.caltech.edu/} \citep{Akeson+2013}.   
\label{fig:miyazaki}
}
\end{figure*}

In this work, we assume continuous observations of 72 days with a 15 min cadence and Gaussian photometric errors that are consistent with simulated photometric error bars shown in Figure 4 of \citetalias{Penny+2019}. 
Periodic correlated noise is expected to be mainly produced by spacecraft systematics and stellar pressure-driven ($p$-mode) oscillations.
Detailed photometric simulations on all systematic errors are computationally expensive.
\citetalias{Penny+2019} applied sub-optimal aperture photometry in their simulated photometric pipeline and added a Gaussian systematic error floor of $1\;{\rm mmag}$ in quadrature into their photometric results to compensate for un-modeled systematic errors.
Timescales of xallarap signals, which are typically $>0.1$ day, are much longer than the expected timescale of p-mode oscillations of main-sequence stars \citep{Broomhall+2009}.
Moreover, \citet{Gilliland+2015} found that most {\it Kepler} stars have a median photometric variability of $\sim0.2$ mmag, with timescale of $<0.6$ hr. 
This is somewhat smaller than \citetalias{Penny+2019}'s error floor of 1 mmag.  

We also assume the source mass and distance to be $M_S=1\;M_{\odot}$ and $D_S=8\;{\rm kpc}$, respectively, and the angular Einstein radius to be $\theta_{\rm E}=0.3\;{\rm mas}$, which approximately leads to 
\begin{eqnarray}
\xi_{\rm E}\sim 4.6\times10^{-5} \left(\frac{M_P}{M_{\rm Jup}}\right)\left(\frac{P_{\xi}}{\rm day}\right)^{2/3}\label{eq:tmp_xa}.
\end{eqnarray}
In our analysis, we consider the orbital inclination of $\lambda_{\xi}=45^\circ$ and $t_0$ to be at center  of 72 days {\it Roman} observing window. 
Our result is hardly dependent on $\theta$ because we adopt $\sigma_{\xi_{\rm E,min}}$ that is independent of the event geometry.
Here we set $\theta=45^\circ$.
For each \textit{W146$_S$}, we adopt the value of the blend flux ratio $\nu$ to be the same with the median value of the $\nu$ distribution that is obtained by the \citetalias{Penny+2019} simulation.
Note that we assume the microlens parallax effect does not affect the measurement of xallarap effect because the xallarap period we focus here is much shorter than 365 days for the parallax, and is thus likely distinguishable.
We do not consider the finite source effect \citep{Witt+1994} because the effect can be easily modeled and is distinguishable from xallarap.
We also do not consider binary-lens events. 
Although the binary-lens event would be more sensitive to the xallarap effect than the single-lens event, it is outside the scope of this work.
We also ignore any accelerations induced by \textit{Roman}'s orbit around L2.

The nominal expected fractional error on the companion mass can be derived from Equation (\ref{eq:xa2}) as
\begin{eqnarray}\label{eq:MP_error}
\frac{\sigma_{M_P}}{M_P} = \sqrt{\left(\frac{\sigma_{\xi_{\rm E,min}}}{\xi_{\rm E}}\right)^2 + \frac{4}{9}\left(\frac{\sigma_{P_{\xi}}}{P_{\xi}}\right)^2},
\end{eqnarray}
by assuming the uncertainties on $\theta_{\rm E}$, $D_S$, and $M_S$ are negligible.
We set the detection threshold for the xallarap companions as $\left(\sigma_{M_P}/{M_P}\right)=0.3$ in the following analysis.
The impacts of uncertainties of $\theta_{\rm E}$, $D_S$, and $M_S$ on the measurements of companion's masses is discussed in Appendix \ref{sec:error_phys}.

\subsection{Xallarap Light Curves}
Figure \ref{fig:sample_lc} shows two samples of model light curves with and without xallarap effect in the top panels.
The residuals of light curves between that with xallarap and without are shown in the second panels.
The left figure represents an event with $t_{\rm E}=30$ days, $u_0=0.1$, $M_P=2\;M_{\rm Jup}$ and $P_\xi=5\;{\rm days}$.
In this case, the maximum deviations from the standard model is about $0.5\%$ of the source flux, which is comparable to the {\it Roman} photometric noise level for \textit{W146}$\sim20$ mag (see Figure 4 of \citetalias{Penny+2019}).
This indicates that brighter and/or high-magnification events are promising targets for {\it Roman} to detect xallarap features induced by a Jupiter-mass planet around the source star.
In the three bottom panels, we plot the components of integrands of the Fisher matrix $\partial F(t_k)/\partial \zeta_i$ at a given time $t_k$.
These panels imply that the observations during $t_0\pm P_\xi$ are most important to determine the xallarap parameters $(\zeta_i)$ related to the mass of companion $M_P$.
However, note that observations further into the wings continue to add to the precision of the period estimate (see the most bottom panels).

For understanding what parts of light curves are important to constrain the parameters, we derive the parameter uncertainties using the Fisher matrix analysis at each time when a {\it Roman} observation is conducted. 
Figure \ref{fig:cumu_sens} represents the cumulative precisions on $\xi_{\rm E}$ and $P_\xi$ as a function of time for the {\it Roman} light curve with \textit{W146}$_S=18$ mag (corresponding to Figure \ref{fig:sample_lc}).
We found that the parameter uncertainties are gradually constrained with increasing data points from the wing of the light curves. 
Moreover, we also derive the cumulative precision on the parameters as a function of the coverage duration for the light curve peak (Figure \ref{fig:cumu_sens2}).
We found that the resultant parameter precisions strongly depend on how long the light curves cover around the event peaks.   
Figure \ref{fig:cumu_sens} and \ref{fig:cumu_sens2} indicate that the xallarap parameters can be measured with incomplete light curves that cover around $t_0\pm P_\xi$.
This is particularly important for \textit{Roman}, which has a short observing window of $72$ days.

\subsection{Xallarap Sensitivity Map} 
At a given $(t_{\rm E}, u_0,$ \textit{W146}$_S)$, we conducted the Fisher matrix analysis on a grid of points over the ranges of $-2 \leq \log(M_P/M_{\rm Jup}) \leq 2$ and $-1\leq\log(P_\xi/{\rm day})\leq3$ with $20 \times 20$ grid points, respectively.
In Figure \ref{fig:sens18}, we present samples of xallarap sensitivity maps in the mass-orbital period plane for the {\it Roman} event with \textit{W146}$_S=18\;{\rm mag}$.
The color maps in each panel represent the distributions of $\sigma_{M_P}/M_P$ and the black lines correspond to contour lines of our detection threshold of $\left(\sigma_{M_P}/M_P\right)=0.3$.
The row and column of a panel correspond to the labeled values of $u_0$ and $t_{\rm E}$.
For example, in the case of $\log(t_{\rm E})=2,\;\log(u_0)=-1.5$ (bottom right panel), the {\it Roman} light curve has a potential to precisely measure the mass of a warm Jupiter with an orbital period of 10-30 days via xallarap. 
We found that $M_P$ is well constrained when $t_{\rm E}$ is longer and $u_0$ is smaller in a given $(M_P,P_\xi)$ grid. 
One also can find that there are sharp cut-offs of the sensitivity with the orbital period of a few dozens days.
This might be because the {\it Roman} observing window of 72 days could not cover the full orbital period of the events beyond the cut-off and thus is insufficient to constrain the xallarap parameters\footnote{When we derive the sensitivity maps, we consider only an observing window of a single season. The sensitivity might extend toward the longer orbital period if we consider all the observing windows. However, it is not expected to be so much because of the results of Figure \ref{fig:cumu_sens2} and because there are long time-gaps between the \textit{Roman} observing windows. Here we focus on only short-period planets and brown dwarfs.}.
This can be expected from the results of Figure \ref{fig:cumu_sens} and \ref{fig:cumu_sens2}.

\section{Prediction of the Yields of Close-in Exoplanets with Xallarap\label{sec:LCS}}
\begin{deluxetable}{lr} [ht!]
\tablecaption{Parameters for {\it Roman} Galactic Exoplanet Survey\label{tab:table}}
\startdata
\\
Survey Area & 1.97 deg$^2$  \\
Mission Baseline & 4.5 years  \\
Seasons & 6 $\times$ 72 days \\
Observation Fields & 7 \\
Microlensing Events with $|u_0|<1$ & $\sim 27000$  \\
\textit{W146} Exposures & $\sim41,000$ per fields \\
\textit{W146} Cadence & 15 minutes \\
Photometric Precision & $\sim$0.01 mag $@$ \textit{W146}$\sim21.15$\\
\enddata
\tablecomments{
The parameters for the Cycle 7 design we use are fully described in \citet{Penny+2019} and \citet{Johnson+2020}.
In this paper, we do not consider any observations with the Z087 filter and other filters that are to be conducted during the {\it Roman} microlensing survey, which could improve our prediction of planet yields to some extent.
}
\end{deluxetable}

\subsection{Simulating on the {\it Roman} Observation}
In this section, we estimate the detection number of close-in planets and brown dwarfs (BDs) in source systems via xallarap during the {\it Roman} mission. 
To simulate the {\it Roman} microlensing survey, we employ the single stellar lens module of the GULLS microlensing simulator \citep{Penny+2013,Penny+2019} which uses version 1106 of the Besan\c{c}on Galactic population synthesis model \citep{Robin+2003,Robin+2012} to generate pairs of lens and source stars.
In Table \ref{tab:table}, we summarize the survey parameters for the Cycle 7 design that we use in our simulation.
The full survey details are described in \citetalias{Penny+2019} and \citet{Johnson+2020}.
Note that we consider only single-lens events whose peaks are within the {\it Roman} observing window.

We classified the simulated {\it Roman} events from GULLS by the values of $(u_0,t_{\rm E},$\textit{W146}$_S)$ into bin $7\times10\times10$ over the ranges of $14<$\textit{W146}$_S<28$ mag, $-2<\log u_0<0$, and $0<\log(t_{\rm E}/{\rm day})<2.5$, respectively.
We generated  the sensitivity maps with the parameters at the center of each bin to be used for all events in each bin.
Then we counted the number of detected events by using the corresponding sensitivity maps at each $(M_P, P_\xi)$ grid.
Figure \ref{fig:miyazaki} shows the resultant detections, i.e., the expected planet yields during the {\it Roman} survey mission if all source stars were to have a planet at each $(M_P, P_{\xi})$ grid point.
The black solid lines represent the contours of the planet yields for $1,10,100$ and $1000$.
The detection sensitivity peaks around $P_{\xi}=20\sim30$ days and there it reaches to sub-Jovian or Saturn masses.
In Figure \ref{fig:miyazaki}, we also plotted observed exoplanets (open dots) from the \dataset[NASA Exoplanet Archive]{https://exoplanetarchive.ipac.caltech.edu/} \citep{Akeson+2013}.
\textit{Roman}'s sensitivity to planets via the xallarap effect largely covers the parameter spaces of hot and warm Jupiters with $M_P>0.5\;M_{\rm Jup}$ and $0.1<P<100$ days, which suggests that this method could be useful to probe the hot and warm Jupiter populations in the Galactic bulge.
Note that we used only a single 72 days season for each event. 
\subsection{Planet Yields}
\begin{figure*}[ht!]
\centering
\includegraphics[scale=0.45,bb=0.000000 0.000000 1057.780191 598.899375]{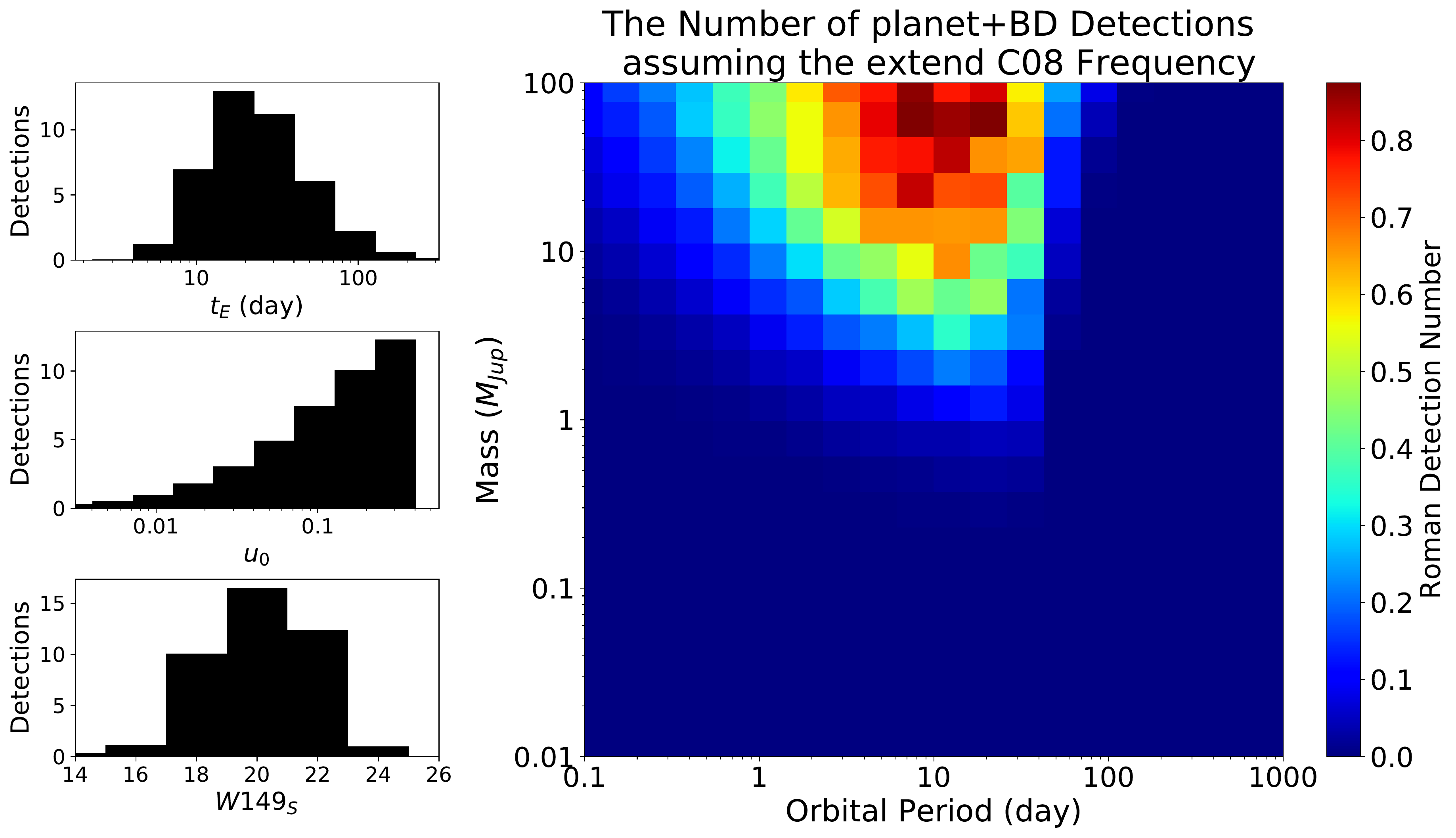}
\caption{
Expected number of the planet and brown dwarf yields during the {\it Roman} mission assuming the extended \citetalias{Cumming+2008} frequency.
The left three panels are the histograms of the yields binned by $t_{\rm E}$, $u_0$, and \textit{W146}$_S$.
The color map in the right panel represents the yields per a given mass-orbital period grid.
\label{fig:summary}
}
\end{figure*}
In order to estimate the planet yields, we assume the secondary mass and period distributions.
We describe the distribution function $f$ as a double power law,
\begin{eqnarray}
\frac{\partial^2 f}{\partial \ln M\;\partial \ln P} = C_{\rm norm}\left(\frac{M_P}{M_{\rm Jup}}\right)^{\alpha_M} \left(\frac{P}{\rm day}\right)^{\beta_P},
\end{eqnarray} 
where $C_{\rm norm}$ is a normalization factor.
In this work, we adopted the power law indexes of $\alpha_M=-0.31\pm0.2$ and $\beta_P=0.26\pm0.1$ derived by \citealp[][(hereafter \citetalias{Cumming+2008})]{Cumming+2008}. 
These values are derived by using 48 RV-detected planets ranging $0.3<M_P<10\;M_{\rm Jup}$ and $2<P<2000$ days that are around Sun-like stars.
We adopted $C_{\rm norm}=0.036\;{\rm dex}^{-2}{\rm star}^{-1}$ to be consistent with a planet frequency of 10.5\% around Sun-like stars in these ranges derived by \citetalias{Cumming+2008}.
Note that we simply extrapolate this \citetalias{Cumming+2008}'s power-law to the ranges $0.01<M_P<100\;M_{\rm Jup}$ and $0.1<P<1000$ days because it is still uncertain.
The extrapolation below $0.3\;M_{\rm Jup}$ hardly affects the final result because the sensitivities to low-mass planets with $< 0.3\;M_{\rm Jup}$ is very low and the \textit{Kepler} survey suggested that the occurrence rate does not significantly rise until below Neptune size of $\sim5\;M_{\oplus}$ \citep[e.g.][]{Fressin+2013}. 
The extrapolations to other range need cautions as discussed below.
We also estimate the yields assuming a simple frequency model of $(\alpha_M,\beta_P)=(0,0)$ and $C_{\rm norm}=0.208\;{\rm dex}^{-2}{\rm star}^{-1}$, which corresponds to single planet per star over the ranges.
This can be considered as the reference yields.

\begin{deluxetable}{lccc} [ht!]
\tablecaption{Expected Yields\label{tab:table2}}
\tablehead{
\multicolumn{1}{c}{} & \multicolumn{2}{c}{Distribution Function} \\
\multicolumn{1}{c}{} & \multicolumn{1}{c}{Simple Model\tablenotemark{a}} & \multicolumn{1}{c}{Extend \citetalias{Cumming+2008}\tablenotemark{b}}
}
\startdata
Companion Mass $M_P\;(M_{\rm Jup})$ & &\\
$10 \leq M_P<100$  & 409.6 & 30.9\\ 
$1 \leq M_P<10$ & 63.3 & 10.1\\
$0.1\leq M_P<1$ & 1.05 & 0.37\\
$M_P<0.1$  & 0.0022 & 0.0014\\ \hline 
Orbital Period $P\;({\rm days})$ \\
$100 \leq P<1000$  & 0.043 & 0.007\\ 
$10 \leq P<100$  & 112 & 15.7\\ 
$1 \leq P<10$  & 243 & 20.7\\ 
$P<1$  & 118 & 5.1\\ \hline \hline
Total & 474 & 42\\
\enddata
\tablenotetext{a}{Simple power-law function with $(\alpha_M,\beta_P)=(0,0)$.}
\tablenotetext{b}{\citet{Cumming+2008} power-law function extended over ranges of $0.01<M_P<100\;M_{\rm Jup}$ and $0.1<P<1000\;{\rm days}$.}
\end{deluxetable}

Figure \ref{fig:summary} represents the expected yields of the {\it Roman} observations assuming the extended \citetalias{Cumming+2008} distribution.
The three left panels show the distributions of $t_{\rm E}$, $u_0$, and \textit{W146}$_S$ for events in which the planet/BD companion around the source is detected.
The yields are expected to largely come from the events with $18<$\textit{W146}$_S<22$ mag, which mostly consists of main sequence source stars. 
Note that although the histogram with $u_0$ indicates that the detections increase towards larger $u_0$, as shown in Figure \ref{fig:sens18}, the events with large $u_0$ are only sensitive to massive companions.
The right panel in Figure \ref{fig:summary} shows the distribution for the number of planets and BDs detections over the parameter spaces of masses and periods assuming the frequency from \citetalias{Cumming+2008}.
Table \ref{tab:table2} summarizes the expected yields with assuming the two different distribution functions. 
Adopting the extended \citetalias{Cumming+2008} distribution, we found that $\sim10$ planets with $M\leq10\;M_{\rm Jup}$ would be detected by xallarap. 
We can expect $\sim30$ companions with $10<M_P<100\;M_{\rm Jup}$ if the extrapolation of \citetalias{Cumming+2008} is correct.
However, due to ``brown dwarf desert'' \citep{Grether+2006}, BD discoveries may by much less common than predicted using the \citetalias{Cumming+2008} frequencies
\footnote{
Under the extend \citetalias{Cumming+2008} function, the existence frequency of BDs around solar-type star is $\sim3\%$ per star.
\citet{Grether+2006} reported that is $<1\%$ per star.
More recently, \citet{Santerne+2016} reported the occurrence rate of BDs within 200 days of the orbital period with $0.29\pm0.17\%$ in the {\it Kepler} transit candidates.
}.
We can test the ``brown dwarf desert'' in the Galactic bulge by applying this method to the upcoming \textit{Roman} light curves.

\section{Discussion \label{sec:DIS}}
\subsection{How to Distinguish Lens Orbital Motion}
If the lensing body is in a binary system, lens orbital motion (LOM) will provide a similar effect to that of xallarap, which has been pointed out in several papers \citep{Rahvar+2009, Penny+2011}.
In a planetary system with orbital period of $P\leq30$ days, a projected angular separation between a lensing host star and its planet in units of $\theta_{\rm E}$, $s=a/\theta_{\rm E}$, is expected to be an order of $s\leq0.06$.
It is unlikely that a lens companion with such a small $s$ provides noticeable light curve deviations by their extremely small caustics. 
For example, \citet{Penny+2011} found that periodic (caustic) features of light curve due to LOM would be most detectable for binary-lens with semimajor axes of $\sim1\;{\rm au}$.
Thus it is difficult to distinguish between xallarap and LOM. 
However the degeneracy may be resolved when following additional high-order effects are observed in the light curves.
The following effects can be observed only in the xallarap events, not in the LOM events.

\begin{description}
\item[Magnified Planet Flux]
Planets orbiting source stars produce reflect light from their host and/or emit their own flux from thermal emission. 
If flux from these companions are magnified, we can observe these contributions in the light curve, as a so-called binary source microlensing event. \citep{Graff+2000, Sajadian+2010}.
Typical flux ratios between solar-type stars and hot Jupiters in the \textit{W146} band is on the order of $\sim10^{-3}$ to $10^{-4}$.
Using this detection channel, \citet{Bagheri+2019} estimated that {\it Roman} will discover $\sim70$ exoplanets from single-lens events and $\sim3$ exoplanets from binary-lens events.\footnote{Note that their most detection samples were composed of hot Jupiters with $a < 0.05$ au, and they assumed a source star has an exoplanet per event.}
We expect that the binary source effect might be observed in the {\it Roman} xallarap single-lens events of $P<5\;{\rm days}$, which corresponds to $\sim50\%$ of the total yields.
It would also help us to constrain the orbital parameters of source systems because it provides the geometric relation between host and planet at the time when the planet is magnified.
Full binary-source modeling including xallarap effect \citep{Miyazaki+2020} is more realistic and might have more sensitivity, but this is out of the scope of this work.
\item[Transiting Source Stars]
If the planet transits the source star, we can also simultaneously observe the transiting signal during the microlensing event \citep{Lewis2001,Rybicki+2014}. 
Typical amplitudes of the transit signal for Jupiter-size planet would be $\sim1\%$ of the source brightness, which will be easily detectable for most xallarap planetary events. 
For example, \textit{Roman} is expected to detect thousands on transiting hot Jupiters, including in the galactic bulge \citep{McDonald+2014, Montet+2017} 
The geometric transit probability of warm Jupiters is $\sim5\%$ so that $\sim5\%$ of xallarap planetary events could be distinguishable from LOM events by the transit signals. 
With the measurement of the planet radius by the transit, we can estimate the density of the planet in the combination with the mass measurement by xallarap.
And, this potentially can test how chemistry affects giant planet structures \citep[e.g.][]{Cabral+2019}.
\item[Ellipsoidal Variations]
Ellipsoidal variations can be caused by tidal effects on the source from the companion \citep[][]{Morris1985}.
The amplitude of the ellipsoidal variation is approximated by 
\begin{eqnarray}
A_{\rm ellip} &\simeq& \alpha_{\rm ellip}\frac{M_P\;\sin \lambda_\xi}{M_S}\left(\frac{R_S}{a_{SC}}\right)^3 \sin \lambda_\xi \nonumber\\
&=& 13\;{\rm ppm}\;\left(\frac{M_P \sin \lambda_\xi}{M_{\rm Jup}}\right)\;\left(\frac{R_S}{R_\odot}\right)^3\nonumber\\
&&\times\left(\frac{M_S}{M_\odot}\right)^{-2}\left(\frac{P_\xi}{\rm day}\right)^{-2}\alpha_{\rm ellip} \sin \lambda_\xi , 
\end{eqnarray}
where $R_S$ is the radius of the source star.
The coefficient $\alpha_{\rm ellip}$ accounts for the stellar limb darkening and gravity darkening: 
\begin{eqnarray}
\alpha_{\rm ellip} = 0.15 \frac{(15+u)(1+g)}{3-u}, 
\end{eqnarray}
where $g$ and $u$ are the coefficients of the gravity darkening and linear limb darkening, respectively \citep{Shporer2017}. 
It would be difficult to observe this effect for the xallarap planetary events.
However, it might be possible for events of substellar source companions.  
\item[Doppler Beaming]
It is known that a line of sight motion due to the orbit causes a periodic variation in the light curve, also known as Doppler beaming \citep{Loeb+2003, Shporer2017}.
The photometric amplitude of the beaming effect $A_{\rm beam}$ is described as below,
\begin{eqnarray}
A_{\rm beam} &=& 2.8\times10^{-3}\;\alpha_{\rm beam}\left(\frac{P_\xi}{\rm day}\right)^{-1/3} \nonumber \\
&\times& \left(\frac{M_S+M_P}{M_\odot}\right)^{-2/3}\left(\frac{M_P \sin \lambda_{\xi}}{M_\odot}\right),
\end{eqnarray}
where $\alpha_{\rm beam}$ is the coefficient accounting of variation of photon amounts in a specific band, e.g. $\alpha_{\rm beam}\equiv1$ in bolometric light.
For longer orbital period of $P_\xi>10$ days, the amplitude of this effect is much stronger than that of the ellipsoidal variations and it is more promising to observe.
\end{description}
These effects may be able to resolve the degeneracy between the xallarap and LOM interpretations in some fraction of xallarap events and provide additional information for source systems.

\subsection{Short-period Populations in the Galactic Bulge Revealed by {\it Roman}}
Some observational results have suggested that there are possible differences in planetary populations between the local neighborhood and the distant region of our Milky Way. 
RV surveys around the local region close to the Sun indicates an occurrence rate of hot Jupiters of order 1\% \citep[][]{Cumming+2008}.  
On the other hand, \textit{Kepler} transit survey toward the Cygnus region suggested the occurrence rate of $0.4\pm0.1\%$ \citep{Howard+2012}, which is approximately half of that in the local neighborhood.
\citet{Penny+2016} used a sample of 31 microlensing exoplanetary systems and suggested the abundance of planets might be less in the Galactic bulge than the disk. 
Such studies would be very important to help understanding whether the planetary formations could depend on its surrounding environment in our Galaxy .

\citet{Montet+2017} predicted that {\it Roman} is expected to discover $\sim100,000$ transiting planets and enable us to directly confirm several thousands hot Jupiters via its secondary eclipses in the light curves.
A large fraction of the transiting planets will belong to the Galactic bulge.
However, in general, the most of host stars are too faint to conduct follow-up RV observations for constraining their planetary mass and avoiding false positives.
\citet{Montet+2017} also suggested some feasibilities for the validation for transiting planets and the estimation of their masses by using phase curve variations although it requires high photometric precisions to observe.
Applying our xallarap method to the {\it Roman} microlensing light curves, we can expect to discover some dozens of hot and warm Jupiters and close-in BDs with measured masses.
These samples will help our understanding of the exoplanet demographics at short orbital periods in the Galactic bulge.
For larger masses, they are also useful to probe the``brown dwarf desert'' around main-sequence stars and study of stellar binary distribution in the Galactic bulge.

\software{{\tt numpy} \citep{Walt+2011}, $\;${\tt matplotlib} \citep{Hunter2007}}
 
\acknowledgments
We would appreciate B. Scott Gaudi, Daisuke Suzuki, Kento Masuda, Akihiko Fukui, Yuki Hirao, Iona Kondo, and Cl\`{e}ment Ranc for valuable comments and discussions.

\appendix
\section{$\Delta\chi^2$ Threshold and Fisher Matrix Analysis \label{sec:deltachi2}}
\begin{figure*}[h!]
\centering
\includegraphics[scale=0.53,bb=0.000000 0.000000 1010.150126 374.022125]{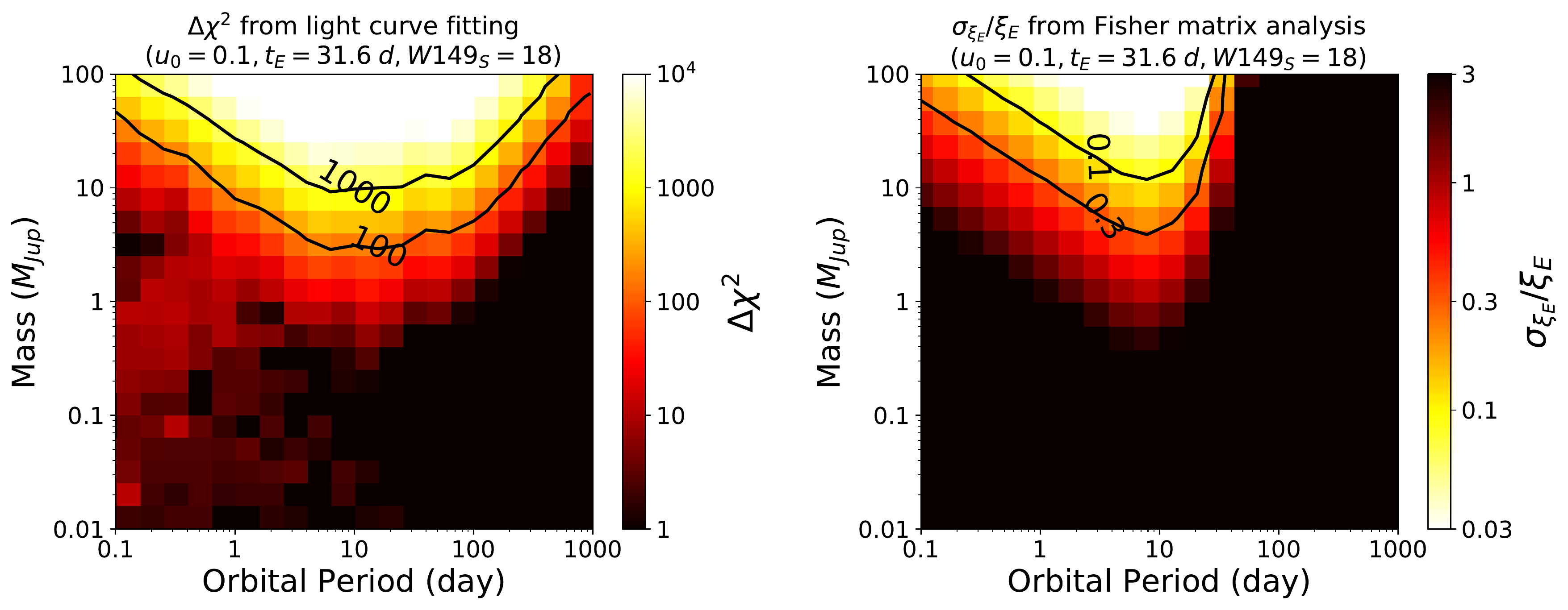}
\caption{
The comparison of distributions over the mass-period diagram between $\Delta\chi^{2}$ from the light curve fitting (left panel) and the parameter uncertainty from the Fisher matrix analysis (right panel).
\label{fig:comp}
}
\end{figure*}
$\Delta\chi^2$ has been often chosen as a detection threshold of planets in most microlensing simulations \citep[e.g.][]{Bennett+1996,Penny+2019,Johnson+2020} possibly because quantifying the detectability can not be analytically solved for binary-lens modeling (see \citealp{Penny+2011} for a discussion of the challenges).
\citet{Rahvar+2009} studied xallarap induced by planetary companions also adopted the $\Delta\chi^2$ threshold of $11.07$ for the detection threshold of planets. 
However, we adopt the Fisher matrix analysis that allows us to quantify the ability not only detecting xallarap signals but also characterizing the physical parameters.
For demonstrating this, we conduct both the Fisher matrix analysis and light curve fitting in the same condition and compare them.

Figure \ref{fig:comp} represents an example of the result with a simulated {\it Roman} event of $(u_0,t_{\rm E},$\textit{W146}$_S)=(0.1,31.6\;{\rm days},18)$.
In the left panel, the $\Delta\chi^2$ distribution is shown as a color map over the mass-period diagram.
Here $\Delta\chi^2=\chi^2_{\rm xallarap}-\chi^2_{standard}$ and $\chi^2_i$ value are calculated fitting each model to the simulated light curve. 
The right panel shows the uncertainty of xallarap amplitude $\xi_{\rm E}$ estimated from the Fisher matrix analysis.
In the parameter space of $(P_{\xi}\geq40\;{\rm day},M_P\geq1\;M_{\rm Jup})$, one finds it is not possible to constrain $\xi_{\rm E}$ well though there should be large $\Delta\chi^{2}$ improvements if we fit the light curve by xallarap model.
This can be because the light curve is strongly affected by only the source acceleration of a single direction and thus only one component of the xallarap vector ${\bm \xi}_{\rm E}$ can be constrained.
For constraining xallarap parameters, it would be required to observe the source accelerations during the full time of source orbits. 
However, we note that it might be possible that the periodogram analysis of the residuals from a standard single-lens fit can constrain the orbital period of the planet sufficiently to enable an estimate of the planet mass \citep{Nucita+2014,Giordano+2017}.

\section{Uncertainties on $D_S$, $\theta_{\rm E}$, and $M_S$ \label{sec:error_phys}}
The ability of \textit{Roman} to measure $\theta_{\rm E}$ and $D_S$ has been studied in several papers.
It is expected that most \textit{Roman} events will have their relative lens-source proper motion ${\bm \mu}_{\rm rel}$ measured via direct detection of lens light in the {\it Roman} images, which enables us to measure $\theta_{\rm E}$ (\citealp{Bennett+2010,Bennett+2020}; \citetalias{Penny+2019}; \citealp{Terry+2020}). 
\citet{Bhattacharya+2018} estimated that \textit{Roman} will measure the lens-source separations with less than 10\% precision for most events.
On the other hand, \citet[][]{Gould+2014} analytically showed that events with photometric precisions of $\leq0.01$ mag have chances to provide the measurements of $\theta_{\rm E}$ with $\leq10\%$ precision via astrometric microlensing in space-based microlensing experiments.\footnote{
They also show that source companions with short orbital periods of $P\leq1$ yr hardly affect the measurements of the astrometric signatures.
}
$D_S$ for bright source events can potentially be measured by the direct parallax (astrometry) measurements in the {\it Roman} survey data \citep{Gould+2015}.
Even if not for bright source events, the combination of the three measurements of the microlens parallax $\pi_{\rm E}$, lens flux $F_{L}$, and $\theta_{\rm E}$ allows us to directly determine $D_S$. 
Moreover, $D_S$ can be statistically estimated with $\sim20\%$ precision using priors of standard Galactic model \citep[e.g.][]{Sumi+2011, Bennett+2014}.
For main sequence stars, $M_S$ is expected to be approximately estimated from the source magnitude and color which are obtained by the multi-band \textit{Roman} photometry (we expect it with less than $20\%$ precision using a stellar isochrone model, e.g., \citealp{Bressan+2012}), and it will become more accurate if $D_S$ is also measured.
For evolved source stars, the mass can be constrained via the age distribution of the bulge.
In this paper, we evaluate the \textit{Roman} ability to characterize the physical parameters of the xallarap companions by adopting $({\sigma_{M_P}}/{M_P})$ in Equation (\ref{eq:MP_error}).
Of course, the resultant errors on the companion's masses will become somewhat larger than the 30\% (${\sigma_{M_P}}/{M_P}=0.3$) due to errors on $D_S$, $\theta_{\rm E}$ and $M_S$.
However, we expect the mass measurements would be possible by {\it Roman} with less than 40\% precision in most cases even if all the errors were included.

\bibliographystyle{aasjournal}

\end{document}